

\magnification=\magstep1

\hsize = 30pc
\vsize = 45pc

\vsize=22.5 truecm
\hsize=16.2 truecm

\voffset=1.1truecm

\font\grand=cmbx10 at 14.4truept

\baselineskip=12pt
\vskip 0.5truecm
\rightline{ DIAS-STP-92-09 \break}
\rightline{ UdeM-LPN-TH-92/92 \break}
\vskip 2cm

\def\bs{\baselineskip=15.5pt}
\bs
\tolerance=5000
\parskip=6pt

\centerline
{\grand On the Lagrangian Realization of the WZNW Reductions}

\vskip 1.3truecm
\centerline{Izumi Tsutsui}
\medskip
\centerline{\it Dublin Institute for Advanced Studies}
\centerline{\it 10 Burlington Road, Dublin 4, Ireland}
\vskip 7mm
\centerline{L\'aszl\'o Feh\'er\footnote*{
On leave from
Bolyai Institute of Szeged University, H-6720  Szeged, Hungary.}
}
\medskip
\centerline{\it Laboratoire de Physique Nucl\'eaire}
\centerline{\it Universit\'e de Montr\'eal}
\centerline{\it Montr\'eal, Canada H3C 3J7}

\vskip 4truecm
\centerline{\bf Abstract}
\vskip 0.4truecm

We develop a phase space path-integral approach for deriving
the Lagrangian realization of the models defined by Hamiltonian
reduction of the WZNW theory. We illustrate the uses of the
approach by applying it to the models of non-Abelian chiral bosons,
$W$-algebras and the GKO coset construction, and show that the
well-known Sonnenschein's action, the generalized Toda action and
the gauged WZNW model are precisely the Lagrangian realizations
of those models, respectively.

\vfill\eject

\def\d{\delta}
\def\p{\phi}

\def\ra{\rangle}
\def\pa{\partial}

\def\i{\int d^2x\,}
\def\hb{\hfill\break}
\def\t{{\rm Tr}\,}

\def\fg{{{\d F}\over{\d g}}}
\def\fj{{{\d F}\over{\d J}}}

\def\si{\sigma}
\def\bsi{\bar\sigma}
\def\D{{\cal D}}

In the last few years it has become clear that
the Wess-Zumino-Novikov-Witten (WZNW) theory [1]
is acting as a \lq master theory'
for a large number of interesting conformally invariant
models in $1+1$ dimensions.
A powerful method for extracting
those models out of the WZNW theory
is the method of Hamiltonian reduction, i.e., the
reduction of Hamiltonian systems with symmetry [2].
The Hamiltonian reductions of
the WZNW theory are defined by placing constraints on the
conserved currents generating the ${\rm left}\times {\rm right}$
Kac-Moody (KM) algebras,
whereby the resulting models possess new (reduced)
symmetry algebras realized explicitly in terms of the
constrained currents [3].
For example, one can in this way
obtain field theoretic models of $W$-algebras,
non-linear extensions of the Virasoro algebra
by conformal primary fields [4], realized
as reduced KM algebras.

Once the constraints are given it is in principle
straightforward to find the reduced Hamiltonian
system.  However, it is not obvious how to find
the underlying Lagrangian from the Hamiltonian system,
because the natural variables of the WZNW theory
(and those of the reduced models)
are not the canonical ones, which can be
defined only locally and are also
quite involved when expressed in terms of the conserved currents.
The purpose of this paper is to present --- through examples ---
a direct approach to
the {\it Lagrangian realization} of the reduced
WZNW models.
The crucial point is that the passage to the Lagrangian
becomes quite simple
if one uses the natural, globally
well-defined variables, rather than the local
canonical variables,
for describing the WZNW
phase space which can be identified
as the cotangent bundle of the loop group
endowed with a modified symplectic form [5,6].
We shall illustrate the uses of our approach
by deriving the Lagrangians
of the reduced WZNW models of non-Abelian
chiral bosons and the models of $W$-algebras.
The respective actions will turn out to be the
much-studied Sonnenschein's action [7,8] and the generalized
Toda action [3].  In these two cases one can in fact find
the reduced variables for the Lagrangians, but
in general it is not easy
to choose them from the WZNW variables.
This problem, however, can always be circumvented
by considering a gauged system, i.e., a
gauged WZNW model, if the constraints are first class
and linear in the current.
We shall obtain two types of gauged WZNW models, one of which
yields the Lagrangian realization of reductions by
general first class chiral constraints [3],
and the other leads to the Lagrangian realization of the
Goddard-Kent-Olive (GKO) construction [9].
The latter gauged WZNW model turns out to be
equivalent to the usual gauged WZNW model considered in [10].

The WZNW theory is most frequently defined in the Lagrangian
formalism [1],
but for our purpose we need its Hamiltonian description,
which we shall recall here briefly [5,6].
In general a Hamiltonian system
is specified by manifold $M$,
Poisson bracket $\{\,,\,\}$ and
Hamiltonian $H$.
Using some irreducible matrix representation
of the underlying
finite dimensional simple Lie group $G$ and its Lie algebra
${\cal G}$, the manifold (phase space)
of the WZNW theory can be given as
$M=\{\, (g,J)\,\vert\, g(\sigma) \in G\, , J(\sigma) \in {\cal G}\}$,
where $g(\sigma)$ and $J(\sigma)$
are periodic,
smooth functions of the space variable $\sigma$.
On the phase space the fundamental Poisson brackets are
\baselineskip=12pt
\footnote*
{
Convention:
$\kappa = - {{k}\over{4\pi}}$
with $k$ being the level of the KM algebra.
Prime and dot stand for
derivative with respect to
the space, $\sigma=x^1$, and time, $\tau=x^0$; and we use
$x^\pm ={1\over 2}(x^0\pm x^1)$.
}
\bs
$$
\eqalign{
\{\,g(\si)\,,\, g(\bsi)\,\}    &= 0 \,, \cr
\{\,\t (a\, J(\si)) \,, \, g(\bsi)\,\}
     &= - a\, g(\si) \, \delta(\si - \bsi)\ , \cr
\{\,\t (a\, J(\si)) \,, \, \t (b\, J(\bsi)) \,\}
     &= \t ([a,b]\, J(\si)) \,\d(\si-\bsi)
                + 2\kappa \t (a\, b) \, \d'(\si-\bsi)\ ,
}
\eqno(1)
$$
for $a$, $b \in {\cal G}$.
Defining the \lq right-current',
$$
\tilde J = - g^{-1} J g + 2 \kappa g^{-1} g^\prime\ ,
\eqno(2)
$$
which forms a KM algebra
with the centre opposite in sign to the one of the KM algebra
formed by the \lq left-current' $J$ in (1), we consider
the Hamiltonian
$$
H = {1\over{4\kappa}} \int d\si \t (J^2 + \tilde J^2)\ .
\eqno(3)
$$
This leads to the usual field equation
$$
\dot g=\{\,g\,,\,H\,\}={1\over \kappa}Jg-g^\prime
\qquad\hbox{and}\qquad
\dot J =\{\,J\,,\,H\,\}=J^\prime\ .
\eqno(4)$$

We shall also need the symplectic 2-form of the WZNW theory.
It is known [5,6] to be given by
$$
\omega = \int d\si \, \t \bigl[\d\, (J\, \d g\,g^{-1})
             + \kappa (\d g\,g^{-1})(\d g\,g^{-1})^\prime \bigr],
\eqno(5)
$$
where $\d$ is the functional exterior derivative.
To see that the above Poisson brackets
can indeed be derived from the sympletic 2-form (5),
we first note that for a function $F=F(g,J)$ on phase space a
vector field $X$ acts as
$$
X(F) = (\d F)(X) = \int d\si\, {\rm Tr}
\Bigl({{\d F}\over{\d g}}  \,
       X(g) + {{\d F}\over{\d J}}  \, X(J) \Bigr),
\eqno(6)
$$
where ${{\d F}\over{\d g}}$ and ${{\d F}\over{\d J}}$
are functional derivatives of $F$ with respect to $g$ and $J$.
Then for two vector fields $X$ and $Y$ we have
$$ \eqalign{
\omega(X,Y)={1\over 2}\int
& d\si\,{\rm Tr\,}
   \Bigl( X(J) Y(g)\, g^{-1} - Y(J)X(g)\, g^{-1} \cr
& + J\,[X(g)\, g^{-1} , Y(g)\, g^{-1}]
     + 2\kappa (X(g)\, g^{-1})
          (Y(g)\,g^{-1})^\prime\Bigr)\ .\cr}
\eqno(7)
$$
We associate the
Hamiltonian vector field $X_F$ to the function $F$ by the formula,
$\d F = 2 \omega (\,\cdot\,,\,X_F), $
which of course means
$$
(\d F)(Y)=2\omega(Y,\,X_F)\ ,\qquad {\rm for} \quad \forall \, Y .
\eqno(8)
$$
Comparing the two sides of (8) given in terms
of (6) and (7) and using the
non-degeneracy of \lq$\rm Tr$', i.e., the non-degeneracy of the
2-form $\omega$, we find that $X_F$ operates as
$$
\eqalign{ X_F(g)
&= \fj\,g\,,\cr X_F(J)
&= \Bigl[ \fj \,,\,J\Bigr] +
    2\kappa \Bigl(\fj\Bigr)^\prime - g\,\fg\ . }
\eqno(9)
$$
Then, as usual, the Poisson bracket of two functions
$F=F(g,J)$ and $H(g,J)$ on $M$ is defined to be
$$
\{ F,H\} \equiv X_H(F) = - X_F(H) = 2\omega (X_H, X_F).
\eqno(10)
$$
By using
eqs.(7), (9) one can now easily obtain the Poisson
bracket of any two
functions on $M$.
The fundamental Poisson brackets (1) arise as special
cases of this general formula [6].

In order to set up the
path-integral of this system,
we define the measure $\D J\, \D g$
by using an invariant volume form of the phase space, which is
obtained by taking
a suitable power of the sympletic 2-form $\omega$.  From (5) it
turns out to be quite simple:
$$
\D J\, \D g =  \prod  \delta J\, \delta g\,g^{-1}\ ,
\eqno(11)
$$
where the product is over the group and the space-time.
Another point to be noted is that the sympletic 2-form (5) is
closed but {\it not exact}.  Still, in general
we can construct a first order Lagrangian
if we extend the parameter region of the phase space,
which is normally parametrized along
one-dimensional curves by the time variable $x^0$,
to a two-dimensional
region by introducing a new parameter in such a
way that the extended region
has the original region of time in its boundary [11].
Usually in the WZNW theory
the extended region is
chosen such that, if it is combined with the
space $x^1$ dimension, it becomes a three dimensional manifold
$B_3$ whose boundary is the $1 + 1$ dimensional
space-time itself.
This construction allows us to formally
write down the {\it phase space}
path-integral of the WZNW theory,
$$
Z = \int \D J\, \D g\,
     e^{i (\int \omega - \int dx^0 H)}.
\eqno(12)
$$
    From (5) the integration of (the pull-back of) $\omega$ performed
over the extended region reads
$$
\int \omega
      = \i \t (J - \kappa \pa_1g\,g^{-1})(\pa_0g\,g^{-1})
                 -{\kappa\over 3} \int_{B_3}\t (dg\,g^{-1})^3.
\eqno(13)
$$
Performing the Gaussian integration of
the \lq momentum type' variable $J$ we obtain the
{\it configuration space} path-integral,
$$
Z = \int  \D g\,
     e^{i S_{\rm WZ}(g)},
\eqno(14)
$$
where $S_{\rm WZ}(g)$ is the usual WZNW action
$$
S_{\rm WZ}(g)
      ={\kappa\over 2} \i \t (\pa_+g\,g^{-1})(\pa_-g\,g^{-1})
                 -{\kappa\over 3} \int_{B_3}\t (dg\,g^{-1})^3 ,
\eqno(15)
$$
as expected.
The expression of the phase-space path-integral (12) is the basis
upon which the Hamiltonian reductions of the WZNW theory
are to be implemented in the following.

\vskip 4mm
\noindent
{\bf Non-Abelian chiral bosons}

One of the simplest examples of the
Hamiltonian reduction of the WZNW theory
is that in which
one of the chiral currents, say the left-current $J$,
is entirely constrained to zero leaving the right-current
$\tilde J$ alone.
Naturally, the reduced theory
is chiral and hence it can
serve as a theory of (non-Abelian) chiral bosons.
Let us examine the content of the reduced theory in detail.

We first note that the constraint surface, $M_c\subset M$,
defined by
$$
\t (a\, J(\si)) = 0,
     \qquad {\rm for} \qquad \forall \, a \in {\cal G},
\eqno(16)
$$
is nothing but the loop group $LG$ of $G$.
Since the constraints (16) satisfy
$$
\{\,\t (a\, J(\si)) \,, \, \t (b\, J(\bsi)) \ra \,\}_{\vert M_c}
               =2 \kappa \t (a\, b) \, \d' (\si-\bsi)\ ,
\eqno(17)
$$
they are almost second class,
apart from the set of zero-modes $J_0 = 0$ in the
Fourier expansion (the horizontal subalgebra) which is first
class.  This set of zero modes
generates a {\it gauge symmetry},
and, accordingly,
the {\it reduced phase space} is identified as
$M_{\rm red}=M_c/G = LG/G$.
(The factorization is by the left-action of $G$ generated
by the zero modes.)
To put it differently,
one would need to use some (local) gauge
fixing conditions\footnote*{
The bundle $LG\to LG/G$ is topologically non-trivial [12].
}
$\chi_0 = 0$ in order to reach $M_{\rm red}$ from $M_c$.
The fact that
the left- and the right-currents commute,
$$
\{\,\t (a\, J(\si)) \,, \, \t (b\, \tilde J(\bsi)) \,\} = 0,
\eqno(18)
$$
implies that $\tilde J$,
which on the constraint surface reads
$$
\tilde J = 2\kappa g^{-1}g^\prime \ ,
\eqno(19)
$$
is gauge invariant, and it is also easy to see
that $\tilde J$ provides a
complete set of gauge invariant functions of $g$.
It follows that the KM symmetry algebra
of the right current survives the reduction,
that is the Dirac brackets of the right-current,
$$
\{\,\t (a\, \tilde J(\si)) \,, \, \t (b\, \tilde J(\bsi)) \,\}^*
     = \t ([a,b]\, \tilde J(\si)) \,\d(\si-\bsi)
       - 2\kappa \t (a\, b) \, \d' (\si-\bsi)\,,
\eqno(20)
$$
are the same as the original Poisson brackets.
On account of (19), the reduced phase space
$LG/G$ can be regarded as a submanifold of
the space of the right-current --- it is in fact
one of the coadjoint orbits of the centrally
extended loop group [12].
(One could obtain any coadjoint
orbit as the reduced phase space by constraining
$J$ to other fixed values instead of zero.)

We next notice that the WZNW Hamiltonian $H$
commutes weakly with the constraints defined by
(16), i.e., $\{\,H\,,\,\t (a\, J(\si))\,\}_{\vert M_c}=0$.
This implies that the WZNW dynamics, defined
by (4), leaves $M_c$ invariant and thus it gives a
natural projection on $M_{\rm red}$.
The Hamiltonian $H$ in (3) reduces on $M_c$ to
$$
H_c={1\over 4\kappa}\int d\si\,\t \tilde J^2=
\kappa \int d\si\, \t (g^{-1} g^\prime)^2\,,
\eqno(21)
$$
which can be used to generate the dynamics of
the reduced theory through the Dirac bracket.
We then find that the gauge
invariant object $\tilde J$ obeys
the chiral field equation
$$
\partial_+ \tilde J = 0\,.
\eqno(22)
$$
As for the group valued field $g$,
the WZNW field equation (4) becomes the
chiral equation $\partial_+ g =0$ upon
restriction to $M_c$, but the chiral solutions
can be subjected to arbitrary
time dependent gauge transformations
without changing their physical meaning,
i.e., their projection on $M_{\rm red}=LG/G$.
In short, $g$ is a chiral
field in the reduced theory
up to the gauge freedom inherent in it.

It is clear that the reduced theory inherites
the conformal invariance of the WZNW theory, since
the constraints (16) are invariant under the
conformal transformations generated by
the usual left and right Virasoro densities,
$$
L(\si) = {1 \over {2\kappa}}{\rm Tr} J^2(\si)
\qquad{\rm and}\qquad
\tilde L(\si) = {1 \over {2\kappa}}{\rm Tr}\tilde J^2(\si).
\eqno(23)
$$
Evidently, $\tilde L$, which is
manifestly gauge invariant, generates the conformal
symmetry of the reduced theory
(while $L$ vanishes upon imposing the constraints).
We have therefore shown that the present
reduction gives rise to
a conformally invariant theory of chiral bosons
possessing a chiral field equation as well as
a single KM algebra in a rather trivial way.

Having described the Hamiltonian system of
chiral bosons, let us find the corresponding Lagrangian.
We are going to read off the Lagrangian from
the phase space path-integral of the reduced theory
following the standard prescription [13]
which implements the constraints
in the path-integral (12).
Namely, we define the phase space path-integral
for our constrained WZNW theory by
inserting the $\d$-functions
of the total, second class constraints,
$\p = \{J_n, J_0, \chi_0\}$ ($n \ne 0$),
in (12)
together with
the associated determinant
factors,
$$
Z = \int \D J\, \D g \,\delta (\phi)
     \,{\det}^{1\over2} \vert \{\phi,\phi\}\vert
     \,e^{i (\int \omega - \int dx^0 H)}.
\eqno(24)
$$
More explicitly, the factors inserted read
$$
\d(\p)\,{\det}^{1\over2} \vert \{\phi,\phi\}\vert
= \d(J_n)\d(J_0)\d(\chi_0)\,
{\det} \vert \{J_0,\chi_0\}\vert
\,{\det}^{1\over2} \vert \{J_n,J_m\}\vert.
\eqno(25)
$$
Note that the second determinant factor in (25)
is just a constant on account of (17).
Then the integration of $J$ yields
the reduced path-integral,
$$
Z = \int \D g \,\d(\chi_0) \, \det
\vert\{J_0, \chi_0\}\vert \, e^{iI_{\rm c}(g)},
\eqno(26)
$$
where
$$\eqalign{
I_{\rm c}(g)
&=
   -\kappa \i \t (\pa_+g\,g^{-1})(\pa_1 g \,g^{-1})
   -{\kappa\over 3} \int_{B_3}\t (dg\,g^{-1})^3\cr
&= S_{\rm WZ}(g)
    - {\kappa\over2} \i \t (\pa_+g\, g^{-1})^2 .\cr
}
\eqno(27)
$$
This is exactly the action of chiral bosons proposed by
Sonnenschein [7].
The invariance of the action (27)
under
$$
g(x) \rightarrow g(x) \, R(x^-)
\qquad {\rm and} \qquad
g(x) \rightarrow T(x^0) \, g(x),
\eqno(28)
$$
for $R(x^-)$, $T(x^0) \in G$, reflects
the fact that there exists the aforementioned
gauge symmetry generated by the zero-modes
in addition to the usual right-symmetry which survived
the reduction.
Because of this, the field equation
$
\pa_+(g^{-1}\pa_1 g) = 0
$
(or $\pa_1(\pa_+g\, g^{-1}) = 0$)
admits the solution
$$
g(x) = g_0(x^0)\,g_{\rm R}(x^-),
\eqno(29)
$$
which is not quite chiral.
However,
it is evident that the non-chirality of $g$
can always be eliminated by using the gauge freedom.
The importance of the gauge invariance
was recognized also in [8],
where the action (27) has been derived
by a coherent state path-integral method.
Here we provided another perspective on
the action of non-Abelian chiral bosons (27) by showing
that it naturally arises from
the Hamiltonian reduction of the WZNW theory, whereby
some of its crucial physical properties,
i.e., a chiral field equation,
a single KM algebra, conformal invariance, are transparent.

\vskip 4mm
\noindent
{\bf Models of $W$-algebras}

Another interesting example can be found by those
WZNW reductions which yield
field theoretic models of $W$-algebras.
There is a natural way to associate a
$W$-algebra to each embedding of the Lie
algebra $sl(2)$ into the simple Lie algebras,
and these extended conformal algebras occur as
symmetry algebras of (generalized) Toda theories
(e.g., see [3] and references therein).
It has been shown earlier by using an
intermediate gauged WZNW theory that the
Toda theory may be regarded
as a reduced WZNW theory, belonging
to left-right dual constraints
which reduce
the two chiral KM algebras to chiral $W$-algebras.
Here we wish to derive the
Toda theory directly from the WZNW theory.

For this purpose,
let us consider a non-compact real Lie group $G$ and
choose a set of elements, $\{M_-, M_0, M_+\}$, which forms
an $sl(2)$ subalgebra of ${\cal G}$.
Then the adjoint
action of $M_0$,
${\rm ad}_{M_0} = [M_0,\,\,]$,
provides a grading of ${\cal G}$ by its
eigenvalues, i.e., by the $sl(2)$-spins.
By using this grading,
we can decompose the algebra ${\cal G}$
into the spaces of positive, zero and negative grades,
$$
{\cal G} = {\cal G}_+ + {\cal G}_0 + {\cal G}_-.
\eqno(30)
$$
For simplicity, we here assume that only integral
spins occur in this decomposition.
(This is true for example for the principal $sl(2)$
subalgebra, which is relevant for the standard
$W$-algebras.)
Let us choose some bases $\{\gamma_i\}$
and $\{\tilde \gamma_i\}$
in ${\cal G}_+$ and ${\cal G}_-$, and
consider the reduction defined by the first class constraints,
$$
\phi_i (\si) =
\t \gamma_i ( J(\si) - \kappa M_- ) =0
           \qquad {\rm and}\qquad
\tilde\phi_i(\si) =
\t \tilde\gamma_i (\tilde J(\si) +\kappa M_+) =0.
\eqno(31)
$$
Note that in this formula we have set to unity the dimensional
constants which in principle occur in front of $M_\pm$
to match the mass dimension.
In spite of having the (hidden) mass parameters the
conformal invariance can still be maintained by
adopting the modified Virasoro density,
$$
L_{M_0}(\si) = L(\si)-2{\rm Tr\,}(M_0 J^{\prime}(\si) ) ,
\eqno(32)
$$
and similarly the right Virasoro density,
which commute weakly with the constraints.  Moreover, it is
possible to find a set of
gauge invariant differential polynomials of $J$
consisting of the Virasoro density (32)
and conformal primary fields which form a $W$-algebra.
Accordingly, the reduced theory possesses the
$W$-symmetry, which is larger than the conformal symmetry [3].

As before, we shall derive the reduced theory by starting
with
the phase space path-integral with first class constraints,
$$
Z = \int \D J\, \D g\,
       \delta (\phi)\,\delta(\tilde\phi)\,
       \delta(\chi)\, \delta(\tilde\chi)\,
       \det \vert \{\phi,\chi\} \vert
       \det \vert \{\tilde\phi,\tilde\chi\} \vert
        \, e^{i (\int \omega - \int dx^0 H)},
\eqno(33)
$$
where we have inserted delta-functions of
gauge fixing conditions $\chi$ and $\tilde \chi$
corresponding to $\phi$ and $\tilde \phi$.
In order to describe the theory in terms of reduced variables,
we associate to (30)
a \lq generalized Gauss decomposition'
of the group $G$,
$$
g=g_+\, g_0\, g_-\,, \qquad {\rm with}{\quad}
g_+=e^{\beta_+},\quad
g_0=e^{\beta_0},\quad
g_-=e^{\beta_-},
\eqno(34)
$$
where $\beta_{0,\pm}$ are from the respective
subalgebras in (30).
We restrict ourselves to considering
the \lq big cell' of the phase space
where $g$ is Gauss decomposable.
Then, by using
the gauge transformations generated by the
first class constraints, we can choose the \lq physical gauge',
$$
g = g_+\,g_0\,g_- \rightarrow g_0,
\eqno(35)
$$
in which
the determinant factors in (33) are constants.  Also, in this gauge
the second delta-function can be written as
$$
\d(\tilde\phi_i ) =
  (\det V(g_0))^{-1} \, \d \bigl( \t \tilde\gamma_i (J
   - g_0 M_+ g_0^{-1}) \bigr),
\eqno(36)
$$
where we have defined
$
V_{ij}(g_0)=\t (\gamma_i g_0\tilde\gamma_j g_0^{-1}).
$
The determinant factor appearing in
(36) is cancelled by the factor arising from the measure
${\cal D}g$ computed in the physical gauge, which in fact is
a result ensured by the
construction of the path-integral reduction [13].
We may now perform the $J$-integration by two
steps as follows.  We first decompose $J$ according to (30) as
$$
J(x) = J_+(x) + J_0(x) + J_-(x),
\eqno(37)
$$
and carry out the $J_+$- and $J_-$-integrations.  Then we find
$$
Z = \int \D J_0 \, \D g_0\,
     \,e^{i (\int \omega_0 - \int dx^0 H_0)}\ ,
\eqno(38)
$$
where
$$
H_0 = \int dx^1 \t \bigl[ {1\over{4\kappa}}
     (J_0^2 + \tilde J_0^2)
        + \kappa g_0 M_+ g_0^{-1} M_- \bigr]\ .
\eqno(39)
$$
Here $\tilde J_0$ is defined similarly to (2),
and $\int\omega_0$ is
given exactly by (13) with $g$ and $J$ replaced by $g_0$ and $J_0$,
respectively.
We thus see from the reduced Hamiltonian (39) and the
phase space path-integral (38) that the reduced theory
(i.e., the generalized Toda theory)
is nothing but a WZNW theory based on ${\cal G}_0$ plus
a potential term.
Finally, the integration of the
momentum variable $J_0$ yields the
configuration space path-integral,
$$
Z = \int \D g_0\, e^{I_{\rm Toda}(g_0)}.
\eqno(40)
$$
The action appearing in (40),
$$
I_{\rm Toda}(g_0) = S_{\rm WZ}(g_0) - \kappa \i
    \t (g_0 M_+ g_0^{-1} M_-),
\eqno(41)
$$
is the generalized Toda action [3] we have been after.

As far as the Hamiltonian reduction leading
to a $W$-algebra is concerned,
one does not need to take the constraints
in the dual form as in (31);
one can take the constraints on the left- and
right-currents independently.
One could also generalize the
reduction by considering arbitrary chiral first class
constraints, although the reduced theory
might no longer possess a $W$-algebra.
For those general cases the above procedure to reach the reduced
Lagrangian may not be easily carried out, since the separation of
the variables into reduced and ignorable ones
is difficult in general.
(Note that in the Toda case it is the grading structure and the
dual nature of the constraints which permitted the simple
$J$-integration by the use of the Gauss
decomposition, which also provided us the reduced variables
in the physical gauge.)
However, even in those cases we can at least have
a Lagrangian realization by a gauge theory, that is,
by the gauged WZNW theory.

To see this, let us choose two subalgebras
$\Gamma$ and $\tilde \Gamma$ of ${\cal G}$ independently and
consider the generalized chiral
constraints,
$$
\phi_i (\si) =
\t \gamma_i (J(\si) -\kappa M ) =0
           \qquad {\rm and}\qquad
    \tilde\phi_i(\si) =
\t \tilde\gamma_i (\tilde J(\si) +\kappa
      \tilde M) =0\,,
\eqno(42)
$$
where $\{\gamma_i\}$ and $\{\tilde\gamma_i\}$ are bases from
$\Gamma$ and $\tilde \Gamma$, respectively,
and $M$ and $\tilde M$ are some constant elements of ${\cal G}$.
The first-classness of the constraints requires that
$$
\t (\gamma_i \gamma_j) = 0
\qquad\hbox{and}\qquad
\t (M\, [\gamma_i\,,\,\gamma_j]) = 0 \ ,
\eqno(43)
$$
and a similar relation for the constraints on the right-current.
(The first relation implies that such constraints
are possible only for a non-compact group $G$.)
The phase space path-integral is given similarly as in
(33) and we first exponentiate
$\delta(\phi)$ and $\delta(\tilde\phi)$ by introducing
two independent Lagrange multiplier fields $A_-\in \Gamma$ and
$A_+ \in \tilde \Gamma$.
By choosing appropriate gauge fixing conditions
$\chi$ and $\tilde \chi$ (e.g., the physical gauge defined
analogously to the Toda case)
the remaining determinant factors and $\delta$-functions
can be taken to be
$J$-independent and thus we can integrate $J$ out as usual.
This way we arrive at the the following effective path-integral
$$
Z = \int \D g\, \D A_- \,\D A_+ \,\, \delta(\chi)\,
      \delta(\tilde\chi)\,
      \det \vert \{\phi,\chi\} \vert\,
      \det \vert \{\tilde\phi,\tilde\chi\} \vert
      \, e^{i I_{\rm eff}(g,A_-,A_+)}\ ,
\eqno(44)
$$
where
$$\eqalign{
I_{\rm eff}(g,A_-,A_+)
&= S_{\rm WZ}(g)
     + \kappa \i \t \bigl[ A_-(\pa_+g\, g^{-1}-M)\cr
&\qquad\qquad
     +A_+(g^{-1}\pa_-g-\tilde M) + A_-gA_+g^{-1} \bigr] \ .\cr}
\eqno(45)
$$
The gauged WZNW action (45), which is invariant under
$$
g \rightarrow \alpha\, g\, \tilde\alpha^{-1}, \quad
A_- \rightarrow \alpha\, A_-\, \alpha^{-1}
               + \pa_- \alpha \, \alpha^{-1}, \quad
A_+ \rightarrow \tilde\alpha\, A_+\, \tilde\alpha^{-1}
               + \pa_+ \tilde\alpha \, \tilde\alpha^{-1},
\eqno(46)
$$
for any
$\alpha(x)\in e^\Gamma$, $\tilde\alpha(x)\in e^{\tilde\Gamma}$,
provides a Lagrangian realization of the WZNW reduction
by the general chiral first class constraints, although it is not
given in terms of the reduced variables alone.

\vskip 4mm
\noindent
{\bf GKO coset construction}

For our final example we consider a coset theory
obtained by the GKO construction [9].
The GKO construction
is designed to produce representations of
the Virasoro algebra
based on the coset $G/H$,
where $H$ is a subgroup of
a simple (or semi-simple) group $G$.
If it is to be realized
in the WZNW context, $H$ will be a diagonal subgroup of the
${\rm left}\times {\rm right}$ group $G \times G$.
Let us find the constraints which lead to the
GKO coset construction.
Actually, we need not look far;
the constraints are given by
$$
\phi_i (\si) =
\t \gamma_i (J(\si) + \tilde J(\si) ) = 0,
\qquad {\rm for} \qquad \gamma_i \in {\cal H},
\eqno(47)
$$
where ${\cal H}\subset {\cal G}$ is the Lie algebra of $H$.
These constraints are first class and indeed
generate the symmetry corresponding to the diagonal subgroup $H$.

It is also
easy to find the modified Virasoro density which commutes
(strongly) with the constraints (47),
$$
L^{G/H}(\si) = L(\si) - L^{H}(\si)\ ,
\eqno(48)
$$
where
$L^{H} = {1\over{2\kappa}}\t_{(H)} J^2$
is the usual Sugawara construction of the Virasoro density with the
summation taken only over the subalgebra ${\cal H}$.
Thus the reduced theory is invariant
under the conformal transformations generated by
$L^{G/H}$, which is the Virasoro density
used in the GKO construction.

The Lagrangian realization of
the reduced theory purely in terms of reduced variables is
difficult to find in general, since
the constraints involve both of the left- and right-currents
simultaneously and thus
it is hard to disentangle them to select the reduced variables
for the Lagrangian.
Therefore we again content ourselves with deriving
a gauged WZNW theory, similarly as in the previous example.
As before, we first introduce a
Lagrange multiplier field $A \in {\cal H}$
and write the phase space path-integral as
$$
\eqalignno{
Z &= \int \D J\, \D g \,\delta (\phi)\, \delta(\chi)\,
     \det \vert \{\phi,\chi\} \vert
     \, e^{i (\int \omega - \int dx^0 H)} \cr
  &= \int \D J\, \D g\, \D A \,\, \delta(\chi)\,
     \det \vert \{\phi,\chi\} \vert
     \, e^{i (\int \omega - \int dx^0 H - A\cdot\phi)},
&(49)
}
$$
where $A\cdot\phi = \int dx^2\,\t A(x)\phi(x)$ and
$\chi = 0$ is a
gauge fixing condition.  Then, by the $J$-integration
we find
$$
Z = \int \D g\, \D A \,\, \delta(\chi)\,
    \det \vert \{\phi,\chi\} \vert
    \, e^{i I_{\rm eff}(g,A)}\ ,
\eqno(50)
$$
where
$$
I_{\rm eff}(g,A) = S_{\rm WZ}(g)
     - \kappa \i \t \bigl[ A (\pa_+g\, g^{-1}) - A(g^{-1}\pa_- g)
     + AgAg^{-1} - A^2 \bigr] \ .
\eqno(51)
$$
The action (51) is invariant under the gauge transformation,
$$
g \rightarrow h\, g\, h^{-1}\ , \qquad
A \rightarrow h\, A\, h^{-1}
               + \pa_0 h \, h^{-1}\ ,
\eqno(52)
$$
for $h = h(x^0) \in H$, but it is not invariant under
the gauge transformation for fully space-time dependent
$h(x)$.  However, if one writes $A = A_0$
one finds that the effective action (51) is nothing but the
\lq axial gauge' $A_1 = 0$ version of the gauged WZNW action
considered in [10], which is invariant under the full $h(x)$.
In summary, we have shown
that the action (51) provides the Lagrangian realization
of the GKO construction, and that its gauge invariant content
is the same as that of the (fully-)gauged WZNW
theory of [10].

\vskip 4mm

In this paper we presented a direct approach to the Lagrangian
realizations of the WZNW reductions and thereby derived
Sonnenschein's action of non-Abelian chiral bosons,
the generalized Toda action,
and the two types of gauged WZNW models
implementing the general chiral first class constraints and
the GKO construction.
This approach should be applicable in general to any
WZNW reduction, and could be used to extract further interesting
(known, or unknown) models out of the WZNW theory.

\bigskip\medskip
\noindent
{\bf Acknowledgement.}
L. F. wishes to thank J. Harnad for explaining
him the symplectic form of the WZNW theory,
which has been crucial for the present work.
I. T. is also indebted to K. Harada and
D. McMullan for helpful conversations.
This paper owes much to our
collaboration with L. O'Raifeartaigh,
P. Ruelle and A. Wipf and we wish to thank them, too.

\vfill\eject \centerline{\bf References}

\vskip 0.8truecm

\item{[1]}
E. Witten, {\sl Commun. Math. Phys.} {\bf 92} (1984) 483.

\item{[2]}
V.I. Arnold,
   \lq\lq {\it Mathematical Methods of Classical Mechanics}",
   Springer, Berlin-Heidelberg-New York, 1978; \hb
A.M. Perelomov,
   \lq\lq {\it Integrable Systems of
   Classical Mechanics and Lie Algebras}",
   Birkh\"auser Verlag, Basel-Boston-Berlin, 1990.

\item{[3]}
See, for example,
L. Feh\'er, L. O'Raifeartaigh, P. Ruelle, I. Tsutsui and A. Wipf,
  {\it On the General Structure of
  Hamiltonian Reductions of the WZNW
  Theory}, Dublin preprint DIAS-STP-91-29, to appear in
  {\sl Phys. Rep.}.

\item{[4]}
A.B. Zamolodchikov,
   {\sl Theor. Math. Phys.} {\bf 65} (1986) 1205; \hb
V.A. Fateev and S.L. Lukyanov,
{\sl Int. J. Mod. Phys.} {\bf A3} (1988) 507;     \hb
F.A. Bais, T. Tjin and P. Van Driel,
   {\sl Nucl. Phys.} {\bf B357} (1991) 632;       \hb
P. Bowcock and G.M.T. Watts,
   {\it On the Classification of Quantum $W$-algebras},
   preprint EFI 91-63, DTP-91-63, and references therein.

\item{[5]}
I. Bakas and D. McMullan,
   {\sl Phys. Lett.} {\bf 189B} (1987) 141;  \hb
K. Gaw\c edzki,
   {\sl Commun. Math. Phys.} {\bf 1991} (1991) 201.

\item{[6]}
J. Harnad and B.A. Kupershmidt,
   {\it Symplectic Geometries on $T^*\tilde G$,
   Hamiltonian Group Actions and Integrable
   Systems}, CRM Montreal preprint (1991),
   to appear in {\sl Commun. Math. Phys.}.

\item{[7]}
J. Sonnenschein,
   {\sl Nucl. Phys.} {\bf B309} (1988) 752.

\item{[8]}
P. Salomonson, B.-S. Skagerstam and A. Stern,
   {\sl Phys. Rev. Lett.} {\bf 62} (1989) 1817;  \hb
M. Stone,
   {\sl Phys. Rev. Lett.} {\bf 63} (1989) 731;
   {\sl Nucl. Phys.} {\bf B327} (1989) 399.

\item{[9]}
P. Goddard, A. Kent and D. Olive,
   {\sl Phys. Lett.} {\bf B152} (1985) 88;
   {\sl Commun. Math. Phys.} {\bf 103} (1986) 105.

\item{[10]}
P. Bowcock,
   {\sl Nucl. Phys.} {\bf B316} (1989) 80;       \hb
K. Gaw\c edzki and A. Kupiainen,
   {\sl Nucl. Phys.} {\bf B320} (1989) 625;      \hb
D. Karabali, Q-H. Park, H.J. Schnitzer and Z. Yang,
    {\sl Phys. Lett.} {\bf 216B} (1989) 307.

\item{[11]}
A.P. Balachandran, G. Marmo, B.-S. Skagerstam and A. Stern,
   \lq\lq {\it Gauge Symmetries and Fibre Bundles}",
   Springer-Verlag Lecture Notes in Physics 188,
   (Springer-Verlag, Berlin and Heidelberg, 1983).

\item{[12]}
A. Pressley and G. Segal,
   \lq\lq {\it Loop Groups}", Clanderon Press,
   (Oxford, 1986).

\item{[13]}
L.D. Faddeev,
   {\sl Theor. Math. Phys.} {\bf 1} (1970) 1; \hb
P. Senjanovic,
   {\sl Ann. Phys.} {\bf 100} (1976) 227.

\bye